\documentclass[sigconf, nonacm]{acmart}
\usepackage{tcolorbox}
\usepackage{soul}
\setul{0.5ex}{0.1ex} 
\usepackage{algorithm}
\usepackage{algpseudocode}
\usepackage{array}
\algtext*{EndIf}
\algtext*{EndWhile}
\algtext*{EndFor}
\algtext*{EndProcedure}
\usepackage{xspace}
\usepackage{pifont}
\usepackage{multirow}
\newcommand\vldbdoi{XX.XX/XXX.XX}
\newcommand\vldbpages{XXX-XXX}
\newcommand\vldbvolume{14}
\newcommand\vldbissue{1}
\newcommand\vldbyear{2026}
\newcommand\vldbauthors{\authors}
\newcommand\vldbtitle{\shorttitle} 

\newcommand\vldbpagestyle{plain} 
\def\cbl{\textcolor{black}}

\newcommand{\ie}{\emph{i.e.,}\xspace}

\newcommand{\eg}{{\emph e.g., \xspace}}

\newcommand{\system}{{\tt VeloANN}\xspace}
\newcommand{\fig}{{Fig.}}
\newcommand{\para}[1]{\noindent {\bf #1} \hspace{2pt}}
\usepackage{enumerate}
\usepackage{enumitem}

\newcommand{\squishlist}{
	\begin{list}{$\bullet$}
		{ \setlength{\itemsep}{2pt}      \setlength{\parsep}{3pt}
			\setlength{\topsep}{3pt}       \setlength{\partopsep}{0pt}
			\setlength{\leftmargin}{5.5mm} \setlength{\labelwidth}{1em}
			\setlength{\labelsep}{0.5em} } }
	\newcommand{\squishend}{
\end{list}  }

\begin{document}
\title{Optimizing SSD-Resident Graph Indexing for High-Throughput Vector Search}


\author{Weichen Zhao}
\authornote{Weichen Zhao and Yuncheng Lu contributed equally to this work.}
\affiliation{%
  \institution{East China Normal University}
}
\email{weichenzhao@stu.ecnu.edu.cn}

\author{Yuncheng Lu}
\authornotemark[1]
\affiliation{%
  \institution{Bytedance}
}
\email{luyuncheng@bytedance.com}

\author{Yao Tian}
\affiliation{%
  \institution{Bytedance}
}
\email{yao.tian@bytedance.com}

\author{Hao Zhang}
\affiliation{%
  \institution{Bytedance}
}
\email{zhanghao.ai@bytedance.com}

\author{Jiehui Li}
\affiliation{%
  \institution{Bytedance}
}
\email{lijiehui@bytedance.com}

\author{Minghao Zhao}
\authornotemark[2]
\affiliation{%
  \institution{East China Normal University}
}
\email{mhzhao@dase.ecnu.edu.cn}

\author{Yakun Li}
\authornote{Corresponding authors.}
\affiliation{%
  \institution{Bytedance}
}
\email{liyakun.hit@bytedance.com}

\author{Weining Qian}
\affiliation{%
  \institution{East China Normal University}
}
\email{wnqian@dase.ecnu.edu.cn}
\begin{abstract}
Graph-based approximate nearest neighbor search (ANNS) methods (e.g., HNSW) have become the de facto state of the art for their high precision and low latency.
To scale beyond main memory, recent out-of-memory ANNS systems leverage SSDs to store large vector indexes.
However, they still suffer from severe CPU underutilization and read amplification (i.e., storage stalls) caused by limited access locality during graph traversal.
We present \system, which mitigates storage stalls through a locality-aware data layout and a coroutine-based asynchronous runtime.
\system utilizes hierarchical compression and affinity-based data placement scheme to co-locate related vectors within the same page, effectively reducing fragmentation and over-fetching.
We further design a record-level buffer pool, where each record groups the neighbors of a vector; by persistently retaining hot records in memory, it eliminates excessive page swapping under constrained memory budgets.
To minimize CPU scheduling overheads during disk I/O interruptions, \system employs a coroutine-based asynchronous runtime for lightweight task scheduling.
On top of this, it incorporates asynchronous prefetching and a beam-aware search strategy to prioritize cached data, ultimately improving overall search efficiency.
Extensive experiments show that \system outperforms state-of-the-art disk-based ANN systems by up to 5.8× in throughput and 3.25× in latency reduction, while achieving 0.92× the throughput of in-memory systems using only 10\% of their memory footprint.

\end{abstract}

\maketitle

\pagestyle{\vldbpagestyle}
\begingroup\small\noindent\raggedright\textbf{PVLDB Reference Format:}\\
\vldbauthors. \vldbtitle. PVLDB, \vldbvolume(\vldbissue): \vldbpages, \vldbyear.\\
\href{https://doi.org/\vldbdoi}{doi:\vldbdoi}
\endgroup
\begingroup
\renewcommand\thefootnote{}\footnote{\noindent
This work is licensed under the Creative Commons BY-NC-ND 4.0 International License. Visit \url{https://creativecommons.org/licenses/by-nc-nd/4.0/} to view a copy of this license. For any use beyond those covered by this license, obtain permission by emailing \href{mailto:info@vldb.org}{info@vldb.org}. Copyright is held by the owner/author(s). Publication rights licensed to the VLDB Endowment. \\
\raggedright Proceedings of the VLDB Endowment, Vol. \vldbvolume, No. \vldbissue\ %
ISSN 2150-8097. \\
\href{https://doi.org/\vldbdoi}{doi:\vldbdoi} \\
}\addtocounter{footnote}{-1}\endgroup


\section{Introduction}
\label{sec:intro}
Approximate nearest neighbor search (ANNS) serves as the core technique behind applications like recommendation \cite{li2018design, li2021embedding, wei2020analyticdb} and retrieval-augmented generation (RAG) \cite{lewis2020retrieval}. 
Among various solutions to the ANN problem, graph-based methods organize vectors into proximity graphs and retrieve neighbors via greedy traversal; these approach have emerged as the de facto state-of-the-art due to their superior accuracy-efficiency balance. 
Nevertheless, existing designs typically assume an in-memory execution model to achieve low latency, which becomes impractical when modern embedding collections grow to billions of vectors.
Therefore, recent efforts turn to out-of-memory ANN systems that seek to scale capacity beyond DRAM while retaining near in-memory performance \cite{fu2017fast, li2018design, wei2020analyticdb}.

To enable out-of-memory (a.k.a larger-than-memory) vector search, systems such as DiskANN~\cite{jayaram2019diskann} and its successors~\cite{wang2024starling, zhong2025vsag, shim2025turbocharging, opensearch2024} adopt a page-based storage layout, where each page holds multiple fixed-size records, and each record contains a vector and its neighbor identifiers. The search process follows the same best-first paradigm as in in-memory graphs, starting from a chosen (or random) entry point and iteratively fetching neighbors from disk until convergence.
To manage memory–disk interaction, production systems adopt two designs. (1) {\em OS-managed caching} via memory-mapped I/O (\eg $\mathtt{mmap}$~\cite{crotty2022you,stoica2013enabling}), used by Qdrant~\cite{qdrant2024} and OpenSearch~\cite{opensearch2024}, offers simplicity but limited control. (2) Self-{\em managed buffer pools} with custom replacement policies, such as in PgVector~\cite{pgvector}, provide explicit control over hot-page residency at the cost of higher complexity (\S 2.4).

\begin{figure}[t]
    \centering
    \includegraphics[width=0.42\textwidth]{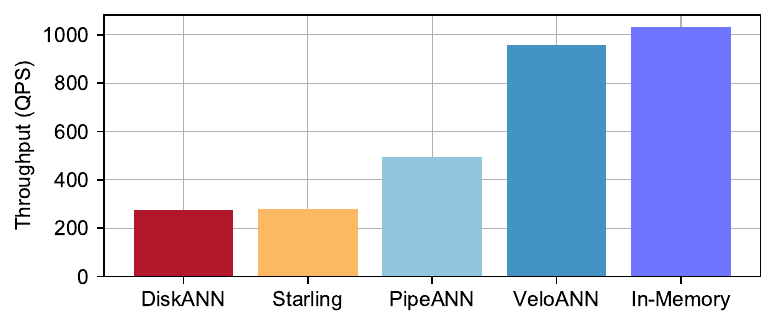}
    \vspace{-2ex}
    \caption{Query throughput comparison on the GIST1M dataset among an in-memory index (Vamana \cite{jayaram2019diskann}), on-disk indexes (DiskANN \cite{jayaram2019diskann}, Starling \cite{wang2024starling}, PipeANN \cite{guo2025achieving}), and our \system. The memory budget is 20\% of the disk index size.}
    \vspace{-13 pt}
    \label{fig:preface}
\end{figure}

In fact, however, {\em current systems simply reuse generic OS or DBMS mechanisms, overlooking the access characteristics of graph-based ANN search, and thus fail to deliver satisfactory performance.}
In such cases, graph traversal exhibits poor spatial locality, causing excessive random I/O and significant CPU underutilization.
As a result, state-of-the-art methods achieve less than half of the query throughput (QPS) of their in-memory counterparts (\fig~\ref{fig:preface}).
As the result of the {\em storage stalls}, performance gap primarily stems from the following factors.

\squishlist
\item \textbf{
The serialized execution of I/O and computation leads to both CPU and I/O underutilized.}
the best-first search algorithm~\cite{guo2025achieving} executes computation and I/O sequentially at each search step, causing excessive CPU idle time while waiting for disk reads, which are typically an order of magnitude slower than computation. In addition, its synchronous I/O execution requires each step to wait for the slowest read in the batch to complete, leaving the I/O pipeline severely underutilized. PipeANN~\cite{guo2025achieving} overlaps the next search iteration with underutilized I/O periods, which reducing query latency but causes a throughput drop due to I/O over-fetching.

\item \textbf{Suboptimal on-disk layouts leads to substantial overprefetching.} Current on-disk layouts typically pack multiple records per page sequentially by record ID. This design suffers from poor data locality, since semantically or spatially related vertices are scattered across different pages.
As a result, an I/O operation often loads an entire page but uses only a single record, leading to severe I/O amplification. It also suffers from page fragmentation, as fixed-size records often leave pages partially filled, wasting storage space and I/O bandwidth. 
Recent disk-based methods \cite{wang2024starling,shim2025turbocharging} adopt graph reordering techniques to place adjacent vertices within the same page  but become impractical at scale due to their high computational overhead.

\item \textbf{Suboptimal buffer management yields low cache hit rates.} Existing paging mechanisms fail to capture the access characteristics of ANN workloads, resulting in low cache hit rate and, consequently, degraded query performance. For instance, on the SIFT1M dataset, when the buffer cache is set to 8\% of the index size, the cache hit rate reaches only 8.49\% for DiskANN~\cite{jayaram2019diskann}.
Even with a generous 20\% buffer allocation, the cache hit rate increases only to 24.4\%, indicating ineffective caching design.
\squishend

\cbl{To tackle these challenges, we design \system, an out-of-memory graph-based ANNS system that eliminates storage stalls through a holistic co-design of compute–I/O scheduling, on-disk data layout, and buffer management. As shown in Figure~\ref{fig:preface}, \system delivers near in-memory performance.} 
To achieve this,
\cbl{\system adopts a coroutine-based asynchronous execution model to fully utilize both CPU and I/O resources. Each query is executed as a coroutine that interleaves distance computation and SSD I/O through non-blocking operations, allowing multiple query coroutines to progress cooperatively. When a coroutine is suspending on an I/O request, the scheduler yields control to other coroutines for computation and resumes it once the I/O completes.
This design effectively overlaps I/O latency with computation, enabling continuous hardware utilization without costly thread-context switches or locking overhead.}

\cbl{To optimize on-disk layouts, \system first quantizes each record to reduce its storage footprint.
The original 32-bit floating-point vectors are quantized into 4-bit-per-dimension codes using ExtRabitQ \cite{gao2024rabitq, gao2025practical}, while the neighbor list is further compressed with efficient integer coding techniques such as delta encoding or Partitioned Elias-Fano coding \cite{ottaviano2014partitioned}.
To support variable-sized record management and fast lookup, \system introduces a new page layout, where the page head stores metadata such as record offsets and IDs, and the variable-sized records are appended from the opposite end. This two-way growth design achieves dense packing to fully utilize available page space. To improve data locality, \system embeds a near-zero-overhead \textit{record coloring} technique into the graph construction process. Upon updating the out-neighbors of each vertex in the initial random graph, \system assigns closely related vertex pairs a shared \textit{color} and places them within the same physical page at the end of construction. By reusing computations already performed during graph building, this design avoids costly graph reordering while achieving the same I/O reduction effect during query traversal. 
 }

 \cbl{To address the low cache efficiency issue, \system redesigns its buffer management from both the data structure and search perspectives.
At the data structure level, it introduces a record-level buffer pool motivated by the highly skewed access patterns in ANN workloads.
For example, on the SIFT1M workload, we found that 47\% of vertices in DiskANN’s graph remain unaccessed during queries, while only 0.1\% of pages are untouched, revealing a clear mismatch between record-level access and page-level caching.
To bridge this gap, \system introduces a record-level buffer pool and employs a lightweight clock-based second-chance policy to pin frequently accessed records in memory and gradually evict cold records. 
}
\cbl{ At the search level, \system proactively fetches promising on-disk candidates ahead of use and develops a cache-aware beam search strategy to prioritize exploring in-memory candidates, achieving fewer I/O stalls and thereby better query performance.}
\cbl{In summary, we make the following contributions:}

\begin{itemize}[leftmargin = 9pt]
    \item \cbl{We design an asynchronous coroutine-based execution framework that dynamically schedules ready queries while others are suspended on I/O, maximizing CPU and SSD utilization.}

    \item \cbl{We propose a novel on-disk page layout that clusters adjacent compressed records into the same page to reduce redundant I/O.}
    
    \item \cbl{We develop a record-level buffer management policy, complemented by asynchronous prefetching and cache-aware beam search to better exploit access skew and improve cache hit rates.}

    \item \cbl{We implement \system\footnote{
   The artifacts will be open-sourced after Bytedance’s review process, which is currently underway.} in Rust with production-grade reliability. Extensive experiments demonstrate that \system delivers up to $5.8\times$ higher QPS and $3.25\times$ lower query latency than state-of-the-art disk-based ANNS systems, while approaching the performance of leading in-memory indexes, \eg using only 10\% of their memory, it achieves $0.92\times$ QPS.}
\end{itemize}

\cbl{The rest of the paper is organized as follows. Section \ref{sec:pre} introduces basic concepts and background. Our \system is detailed in Section \ref{sec:velo} and cache-aware ANNS procedure Section \ref{sec:beam_search} followed by experimental results in Section \ref{sec:exp}. Section \ref{sec:relatedwork} reviews related work, and we conclude in Section \ref{sec:conclusion}.}

\section{Background}
\label{sec:pre}
This section first introduces the fundamentals of ANNS and graph-based indexes for ANNS, then reviews techniques for asynchronous execution and database buffer pool management.

\subsection{Approximate Nearest Neighbor Search}
\cbl{In this paper, we study the approximate nearest neighbor search (ANNS) problem in Euclidean space. 
Let the dataset be $X = \{x_1, \dots,$ $x_n\}$, where each data point $x_i \in \mathbb{R}^d$ is a $d$-dimensional vector. 
Given a query vector $q \in \mathbb{R}^d$, the goal of nearest neighbor search is to find the top-$k$ data points in $X$ that are closest to $q$ under the Euclidean distance metric $\|\cdot, \cdot\|_2$, namely:
\begin{equation}
R = k\text{-}\arg\min_{x_i \in X}\|q - x_i\|_2 .
\end{equation}
Due to the curse of dimensionality \cite{indyk1998approximate}, computing the exact top-$k$ neighbors is prohibitively expensive in high dimensions. Therefore, ANN search is commonly adopted, which trades a bit accuracy for substantial efficiency gains.  For an ANN result $R'$ with $|R'| = k$, the retrieval accuracy is typically measured by \emph{Recall@k}, defined as
\begin{equation}
\text{Recall@k} = \frac{|R \cap R'|}{k}.
\end{equation}}

\subsection{Graph-based ANNS Index}
As reviewed in Section \ref{sec:relatedwork}, numerous graph-based indexes have been proposed for ANN queries \cite{azizi2023elpis, chen2021spann, fu2016efanna, fu2021high, fu2017fast, harwood2016fanng, malkov2018efficient, peng2023efficient, wang2021comprehensive}. These approaches achieve state-of-the-art search performance by efficiently identifying nearest neighbors through graph traversal, typically examining only a limited number of candidate vectors~\cite{munoz2019hierarchical}. Specifically, during the indexing phase, these methods construct a graph structure where each data vector corresponds to a vertex. Upon receiving a query, the system employs a greedy beam search algorithm~\cite{fu2021high,fu2017fast,jayaram2019diskann,malkov2018efficient}. Throughout the search process, the algorithm maintains a beam set with a maximum capacity of $n_b$ vectors. The search initiates by adding an entry point to the beam set and proceeds iteratively. In each iteration, the algorithm identifies the unvisited vector within the beam set that is closest to the query vector, marks it as visited, and computes exact distances between the query vector and all its graph neighbors. These neighbors are subsequently inserted into the beam set. If the beam set exceeds its capacity $n_b$ after insertion, only the $n_b$ nearest vectors are retained. The algorithm terminates when all vectors in the beam set have been marked as visited, at which point the nearest vector in the beam set is returned as the result. It is evident that a larger beam size results in the examination of more vectors during search, consequently yielding higher accuracy at the cost of increased latency.

\subsection{Asynchronous Execution Framework}
Modern I/O-intensive data systems increasingly adopt asynchronous execution models to optimize performance and resource utilization.
This model leverages asynchronous I/O to allow issuing multple I/O requests in the background, so that the thread don't need to block on the synchronous I/O.
Complementary with this, organizing and scheduling tasks as coroutines---functions that can suspend and resume its execution---to effectively overlap computation with I/O operations.

\para{Asynchronous I/O.}
To overcome CPU underutilization in synchronous I/O, where threads block awaiting I/O completion, modern data systems employ asynchronous I/O (AIO).
AIO enables threads to initiate I/O operations and continue execution, receiving notifications upon completion.
This non-blocking approach is fundamental for building high-performance, I/O-intensive systems.
Key implementations include the established POSIX AIO and the more recent Linux \texttt{io\_uring} interface.
POSIX AIO allows applications to initiate one or more I/O operations that are performed in the background.
{\tt io\_uring} works by creating two circular buffers, {\ie submission and completion queues}, to track the submission and completion of I/O requests, respectively. 
Keeping these buffers shared between the kernel and application helps boost the I/O performance by eliminating the need to issue extra and expensive system calls to copy these buffers between the two.
It also enables batched I/O requests to allow multiple I/O operations to be processed concurrently.

\para{Coroutines.} 
Coroutines are special functions that can suspend (yield CPU time) voluntarily and be resumed later by a user-space scheduler as needed. 
Specifically, upon issuing an asynchronous I/O operation, a coroutine can suspend execution, allowing the scheduler to run another ready coroutine; once the I/O completes, it resumes to continue its task.
Stackless coroutines, as standardized in C++20 and Rust, provide a concurrency model with minimal overhead for construction and context switching, often less than the cost of a last-level cache miss~\cite{he2020corobase}.
Unlike traditional coroutines, they do not maintain their own stacks; instead, they execute on the call stack of the runtime scheduler.
While coroutine invocation resembles a standard function call, its state---including local variables that persist across suspension points---is preserved in dynamically allocated memory that survives suspend and resume cycles.

For instance, Rust natively supports asynchronous programming.
It abstracts asynchronous operations with the \texttt{Future} trait, which represents a computation whose result may not yet be available (\eg I/O operations) and makes progress when polled.
With Rust's syntactic sugar (\eg \texttt{await}), asynchronous I/O can be expressed in a sequential style, while a pluggable runtime scheduler ({\ie the scheduling logic can be customized}) schedules and executes other tasks during I/O wait periods.

\subsection{Larger-than-Memory Database Techniques}
\label{sec:buffer_pool_tech}
While in-memory databases offer superior performance, their capacity is limited by the available main memory.
To handle datasets that exceed this capacity, database systems must utilize secondary storage and manage the movement of data pages between memory and disk.
The techniques for managing this data swapping can be broadly categorized into two types:

\para{OS-controlled caching.}
Memory-mapped (\texttt{mmap}) file I/O is an OS-provided feature that maps file contents on secondary storage into a process's virtual address space, allowing the database to access pages via pointers as if they resided entirely in memory. The OS transparently loads pages on-demand upon access and evicts them according to its internal policies when memory pressure arises. Owing to its simplicity and low engineering cost, many vector databases (\eg Qdrant~\cite{qdrant2024}, OpenSearch~\cite{opensearch2024}) employ \texttt{mmap} to handle larger-than-memory datasets. However, as Crotty et al.~\cite{crotty2022you} argue, \texttt{mmap} is often ill-suited for DBMSs because the DBMS loses control over page faulting and eviction, and the OS's virtual memory subsystem can be too slow for modern NVMe SSDs.

\para{Self-Managed Caching.} To gain full control over page loading and eviction, many databases implement their own buffer management in user space.
The traditional approach uses a hash table to map page identifiers (PIDs) to their location in the buffer pool.
Accessing a page thus requires a hash table lookup, and a miss triggers a read from secondary storage.
While flexible, the overhead of these lookups can be substantial.
To mitigate this cost, recent designs employ more memory-efficient techniques.
For instance, LeanStore~\cite{leis2024leanstore} uses pointer swizzling~\cite{graefe2014memory,kemper1995adaptable}, which replaces a page's PID with its direct memory pointer within other data structures.
This allows page accesses on a cache hit to be performed via a direct pointer dereference, bypassing the hash table.
However, this technique requires strict exclusive ownership control ({\ie} a page can only be referenced by a single pointer) to handle concurreny control, making it unsuitable for graph structures where a node can have multiple references.
Another approach, VMCache~\cite{leis2023virtual}, builds on \texttt{mmap} but reclaims control by using the \texttt{madvise} system call to provide caching and eviction hints to the OS, combining the simplicity of \texttt{mmap} with the control of a self-managed cache.

\section{VeloANN Design}
\label{sec:velo}
\begin{figure}
    \centering
    \includegraphics[width=0.475 \textwidth]{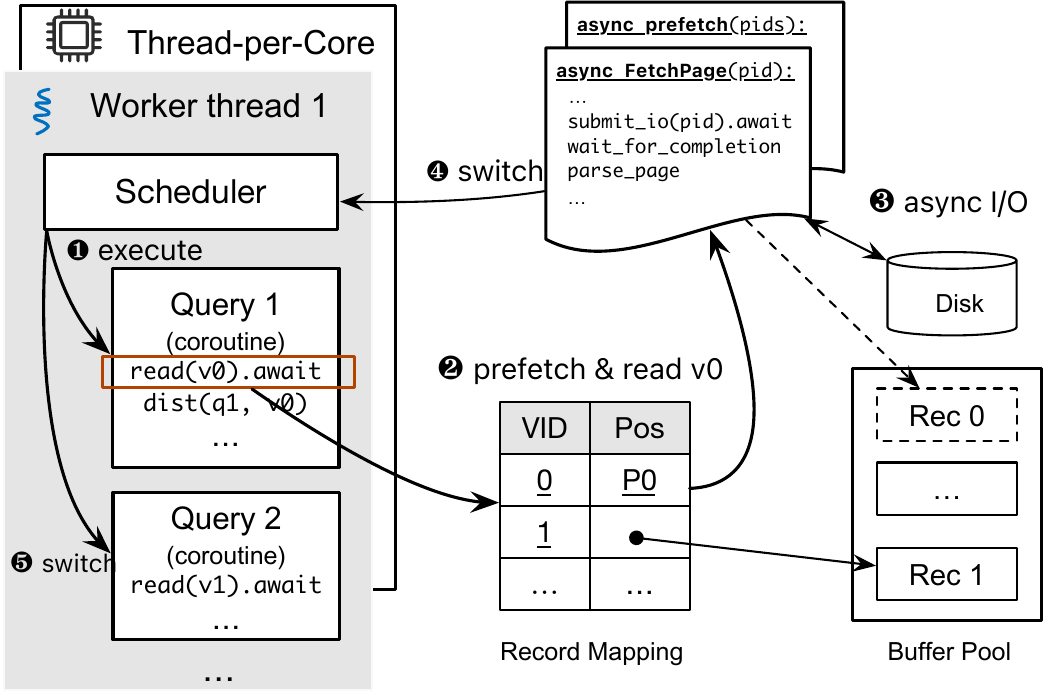}
    \caption{Overview of asynchronous execution in \system.}
    \Description{}
    \label{fig:overview}
\end{figure}

We present \system, a I/O-efficient ANNS system for handling workloads that exceed main-memory capacity.
As shown in \fig~\ref{fig:overview}, \system adopts a thread-per-core asynchronous execution model where each worker thread runs a dedicated scheduler that models incoming batched queries as coroutines. This design enables a worker to interleave multiple queries, hiding I/O latency by switching to a ready coroutine when another awaits data from disk~(\ding{182}). A running coroutine traverses the graph, issuing access requests for records it needs~(\ding{183}). If a requested record is on disk ({\ie} a cache miss), the coroutine initiates an asynchronous I/O operation to fetch it~(\ding{184}) and yields control~(\ding{185}). If the record is already in memory, the coroutine continues its execution. The scheduler then switch the control to next ready ANNS coroutine. The scheduler completes the cycle by polling the I/O completion queue and readying the corresponding coroutine for resumption once its data arrives~(\ding{186}). Through these I/O-hiding mechanisms, \system sustains high CPU utilization by reducing I/O stalls along the query execution path.

This section details the design of \system.
We first introduce its coroutine-based asynchronous execution model, which overlaps computation and I/O~(\S\ref{sec:coroutine_execution}).
We then describe the on-disk page layout (\S\ref{sec:page_layout}) and record-level buffer pool that enable the index to scale beyond main memory capacity~(\S\ref{sec:buffer_pool}).

\subsection{Coroutine-based ANNS Execution}\label{sec:coroutine_execution}
\begin{figure}
    \centering
    \includegraphics[width=0.48 \textwidth]{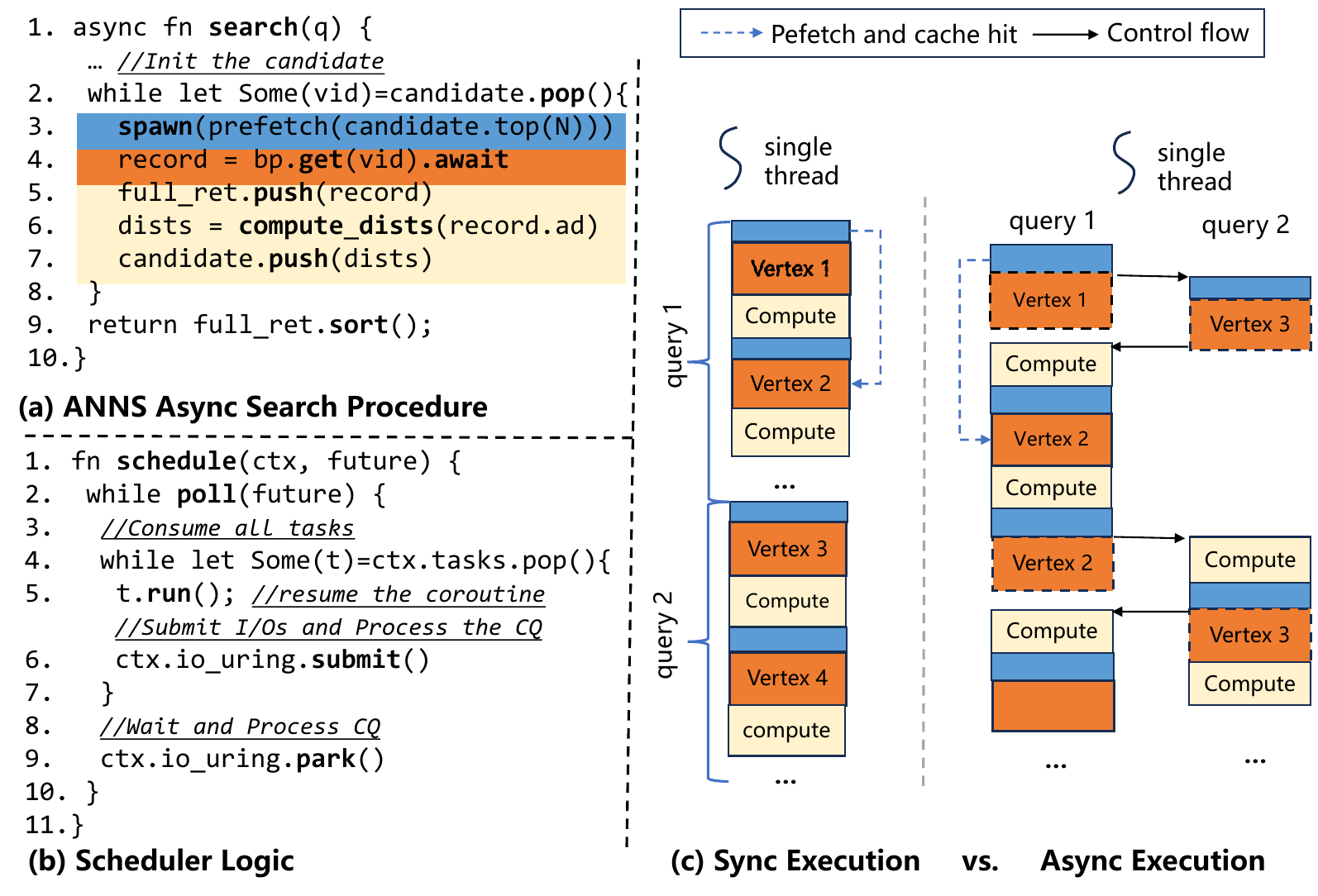}
    \caption{Coroutine-based ANNS Execution. Each thread has its own scheduler that batches incoming queries and interleaves multiple query coroutines.}
    \Description{}
    \label{fig:coroutine_execution}
\end{figure}
Existing disk-based ANNS systems like DiskANN~\cite{jayaram2019diskann} suffer from significant CPU underutilization.
Our measurements show only 57\% CPU utilization on Sift1M, as their sequential compute-I/O model forces the CPU to idle during disk reads, which are an order of magnitude slower than computation.
Although beam search mitigates this by reading multiple records concurrently, the synchronous execution pattern persists.
Each search step remains bottlenecked by the slowest I/O operation, preventing high CPU utilization.
To address this bottleneck, \system introduces a coroutine-based execution model with holistic redesign of the ANNS query processing to hide I/O stalls.
By modeling each query as a coroutine, the scheduler can switch from one awaiting I/O completion to another that is ready for computation.
As illustrated in \fig\ref{fig:coroutine_execution}, this asynchronous query processing approach allows \system to saturate the CPU:

\para{Asynchronous ANNS Procedure.} Panel (a) in \fig\ref{fig:coroutine_execution} illustrates the asynchronous, coroutine-based ANNS procedure.
The search begins by initializing a candidate list with entry points in the graph index.
The asynchronous search procedure then iteratively refines the list using a best-first search strategy with three main phases (lines 2-10):
(i) The coroutine issues non-blocking prefetch requests for the top-$N$ candidate vertices to improve the cache hit rate for subsequent search steps (line 3, detailed in \S\ref{sec:beam_search}).
(ii) It then requests the vertex record (i.e., vector and adjacency list) from the buffer pool (line 4).
If the record is not cached, the coroutine suspends and yields control to the scheduler, awaiting I/O completion (line 5).
(iii) Once the record data is loaded, the scheduler resumes the coroutine, which then computes the distance between the query and the fetched vertex and estimates distances to its neighbors using their quantized vectors to update the candidate list (lines 5-7).
This process repeats until the candidate list is exhausted.

\para{I/O-Aware scheduler.}
As illustrated in Panel (b) of \fig\ref{fig:coroutine_execution}, each worker threads is pinned to a CPU core and runs a dedicated scheduling loop to process incoming ANNS queries in batches.
The scheduler runs a continuous loop util the main function has terminated (line 2).
When the control yields to the scheduler, the scheduler then retrieves ready coroutines from the task queue and resumes their execution (lines 4-5).
When an ANNS coroutine suspends (\eg on a cache miss), the scheduler submits the pending I/O requests via an asynchronous I/O driver (\eg \texttt{io\_uring}) and processes the completion queue from the interface, pushing coroutines with completed I/O back to the task queue (line 6).
If the ready queue is empty, this means all coroutines are awaiting I/O completion.
In this scenario, the scheduler busy-polls the I/O completion queue until a coroutine can be made ready.

The scheduler limits the batch size $B$, {\ie} the number of concurrently executing queries per thread, to balance the trade-off between throughput and latency.
While a larger batch size can improve throughput by enabling the scheduler to more effectively overlap I/O with computation, it may also increase average query latency due to context switching and scheduling overhead.
To optimize this balance, we formulate the batch size as $B = \lceil \alpha \cdot I / T \rceil$.
In this formula, $I$ is the average disk I/O latency, $T$ is the average execution time of the computational task between two I/O events, and $\alpha$ is a hyperparameter representing the expected frequency of I/O events per computational task.
This formulation is derived from the principle of overlapping computation with I/O to keep the CPU utilized.
Since a coroutine computes for an average of $T/\alpha$ before an I/O of latency $I$, the number of coroutines needed to keep the CPU busy during that I/O is $I / (T/\alpha) = \alpha \cdot I / T$.
The value of $\alpha$ is empirically determined and depends on factors like the cache hit ratio and coroutine switching overhead.

\para{Sync~vs.~Async execution.} Panel (c) demonstrates the difference between synchronous and asynchronous ANNS execution.
In the synchronous model, when the CPU processes a query and encounters a random I/O operation (\eg accessing a vertex record not in memory) during graph traversal, a cache miss causes the CPU to stall until the record data is fetched from disk.
In contrast, the asynchronous model allows CPU to issue the request for the required vertex record and immediately switch to process another query coroutine managed by the scheduler.
Each coroutine handles an ANNS query and suspends execution upon an I/O operations.
When the scheduler switches back to the previously suspended coroutine, the CPU does not need to wait for the record as it has been loaded into the in-memory buffer pool.
This overlapping between data fetching and query computation sustains high CPU utilization and significantly improves query throughput.

\begin{table}[t]
\centering
\caption{Cache hit rate (\%) of different page replacement policies with varying buffer ratios on a custom page-level buffer pool for DiskANN.}
\label{tab:cache-hit-rate}
\begin{tabular}{@{}lccccc@{}}
\toprule
& \textbf{10\%} & \textbf{20\%} & \textbf{30\%} & \textbf{40\%} & \textbf{50\%} \\ \midrule
LRU & 13.0\% & 25.8\% & 35.8\% & 45.1\% & 54.0\% \\
FIFO & 15.0\% & 26.0\% & 35.9\% & 45.3\% & 54.3\% \\
Random & 14.9\% & 25.0\% & 34.8\% & 44.1\% & 52.8\% \\ \bottomrule
\end{tabular}
\end{table}

\begin{figure}[t]
    \centering
    \includegraphics[width=0.47\textwidth]{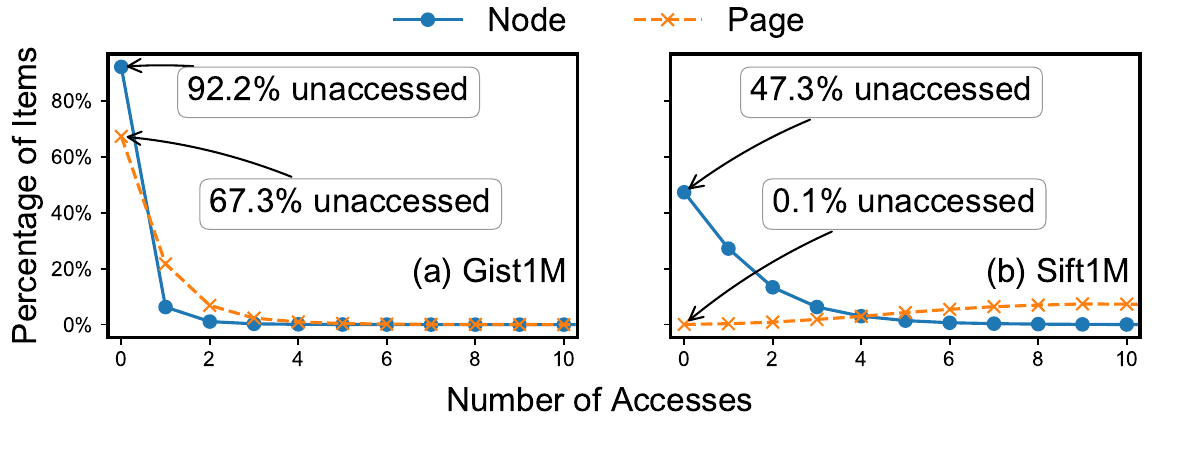}
    \caption{Access frequency distributions at node- and page-level granularities for DiskANN on Sift1M and Gist1M.}
    \label{fig:motivation-access}
\end{figure}

\subsection{Buffer Pool Management}
\label{sec:buffer_pool}
Paging methods, such as buffer pools or {\tt mmap}, are fundamental to database systems but are often ill-suited for ANNS workloads.
They struggle to achieve high cache hit rates, given the unique and often unpredictable memory access patterns of ANNS queries.
To demonstrate this, we experimented on the Sift1M dataset with a page-level buffer pool for DiskANN, evaluating several common cache replacement policies while varying the buffer size as a fraction of the index size.
As shown in Table~\ref{tab:cache-hit-rate}, the cache hit rate improves only linearly with increased buffer allocation.
This low access locality stands in stark contrast to traditional database workloads like TPC-C, where 80\% of accesses typically target just 20\% of the data~\cite{leutenegger1993modeling}.
Interestingly, sophisticated replacement policies (\eg LRU, FIFO) offer only marginal improvements over a simple random replacement strategy, underscoring their limited effectiveness for ANNS workloads.

To address this, our key opportunity lies in {\bf exploiting access skew by decoupling vertex-based records from page-based organization.}
As illustrated in \fig\ref{fig:motivation-access}, we measured the access frequency of each vertex and page after executing a complete workload on the Sift1M and Gist1M datasets.
We observe that while vertex-level statistics reveal a significant access skew, this phenomenon is difficult to exploit with paging methods.
Consequently, accessing a single record necessitates loading its entire page into the cache, even if other records on that page are infrequently accessed.
This reveals a pronounced locality mismatch in Sift1M: 47.3\% of vertices remain unaccessed, yet only 0.1\% of pages are untouched.
A similar disparity is observed in Gist1M, undermining the effectiveness of traditional page-based cache replacement policies.

\para{Record-level Buffer Pool.}
Based on this observation, \system adopts a record-level buffer pool which only caches the requested records and discards the co-resident records on the same page.
However, achieving performance comparable to that of in-memory ANNS systems with a generic buffer pool is challenging for two main reasons.
First, traditional buffer management techniques are ill-suited for graph-structured data; hash table-based methods incur indirection overhead, while pointer swizzling cannot support graph structures, as detailed in \S\ref{sec:buffer_pool_tech}.
Second, a fundamental granularity mismatch exists between the record-based in-memory layout and the page-oriented on-disk layout.

To address these challenges, \system implements a \textbf{slotted buffer pool} to efficiently manage the ANNS graph layout.  It allocates an anonymous virtual memory region sized to a configured fraction of the total index size (\ie the buffer ratio) and partitions it into slots, each sized to hold one vertex-based record.
A free list tracks the available slots.
To overcome the limitations of pointer swizzling for graph structures, \system employs a compact indirection array (the \textit{record mapping array} in \fig\ref{fig:overview}).
Because vertex IDs are assigned contiguously, this array provides $O(1)$ mapping from a vertex ID to a record's physical location and centralizes data-residency management, thereby resolving pointer swizzling's incompatibility with graph structures.
Each entry in this array is a hybrid pointer in which the most significant bit encodes residency: bit 1 indicates in-memory, with the remaining bits indexing a slot in the buffer pool; bit 0 indicates on-disk, with the remaining bits encoding the page ID of the record's location.
This design eliminates access overhead for memory-resident records.

This buffer pool design requires a synchronization mechanism to manage concurrent operations (\eg loading and eviction) in its slots.
To this end, we explicitly maintain a state for each slot, managed via atomic operations.
Since the number of slots is a subset of the total vertex count, the memory footprint of this state management is minimal.
After startup, all slots are initialized to the \texttt{Free} state, and subsequent record access operations check the corresponding state entry before proceeding.
We now introduce the basic operations of the buffer pool to illustrate the state transitions of a slot, as shown in the state diagram in \fig\ref{fig:eviction}:

\begin{figure}[h]
    \centering
    \includegraphics[width=0.4 \textwidth]{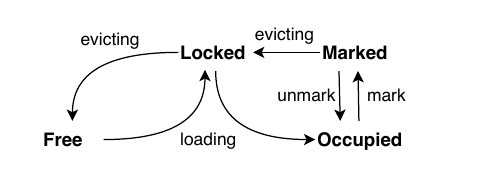}
    \caption{The state transitions of record states.}
    \Description{}
    \label{fig:eviction}
\end{figure}

\para{Loading.}
The loading process is initiated when the \textit{record mapping array} indicates that a requested vertex record resides on disk.
The buffer manager first acquires a slot from the free list; if none is available, an eviction is triggered.
Once a slot is secured, its state is atomically set to \texttt{Locked} via a compare-and-swap (CAS) operation to prevent race conditions.
An I/O request is then dispatched to fetch the required page from disk.
Upon I/O completion, the target record is parsed from the page, loaded into the slot, and the slot's state is updated to \texttt{Occupied}.

\para{Eviction.} When the buffer pool approaches full, an eviction coroutine is invoked to free slots for new records.
As established in our analysis (Table~\ref{tab:cache-hit-rate}), conventional replacement policies are ill-suited for ANNS workloads; consequently, our design prioritizes low eviction overhead over cache effectiveness.
To this end, we implement a clock-based replacement policy with a second-chance mechanism.
This policy employs a clock hand that sweeps over the buffer pool slots in a circular manner.
When the hand encounters a slot in the \texttt{Occupied} state, it transitions it to \texttt{Marked} and advances.
If a query accesses a \texttt{Marked} slot, its state reverts to \texttt{Occupied}, giving it a second chance.
If the hand encounters an already \texttt{Marked} slot, it is selected for eviction.
This process continues until a target number of slots are freed, ensuring that frequently accessed records (\eg entry points) remain resident with minimal overhead.
The eviction process is designed as a scalable, concurrent coroutine to minimize contention with foreground query processing.

\begin{figure}[t]
    \centering
    \includegraphics[width=0.45\textwidth]{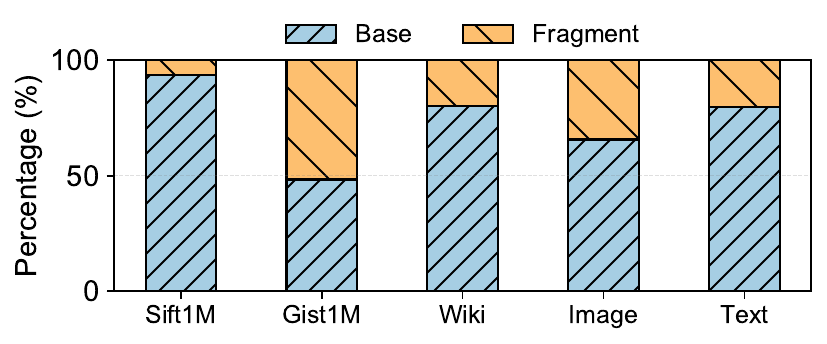}
    \caption{Internal page fragmentation percentage of the on-disk layout for ANNS graph indexes.}
    \label{fig:fragment_percentage}
\end{figure}

\subsection{Physical Index Layout}\label{sec:page_layout}
ANNS graph indexes typically employ a vertex-based layout, storing each vertex's vector and adjacency list together as a single record.
This layout leverages spatial locality in graph traversal, as a vertex's vector and its out-neighbor list are typically accessed together during search.
Each record is packed into fixed-size disk pages.
To ensure uniform record sizes for efficient packing and lookup, each record's adjacency list is padded to a fixed bound $R$.
However, this layout incurs internal page fragmentation: fixed-size records often leave pages partially utilized, wasting space.
As illustrated in \fig~\ref{fig:fragment_percentage}, fragmentation remains low at small vector dimensionalities (\eg $d=128$ for Sift1M).
By contrast, as dimensionality increases, the upper bound on internal fragmentation rises; \eg Gist1M reaches up to 52\%, while other high-dimensional datasets are around 25\%.
Large, fixed-size records induce padding, leading to partial page utilization.
These observations motivate compressing vertex records and adopting a more compact on-disk layout to improve page packing efficiency and reduce internal fragmentation.

\para{Compressed Vertex-Based Record.}
We minimize record size by compressing both the vector and adjacency list of each vertex record.
Vectors are hierarchically compressed (\eg ExtRaBitQ~\cite{gao2024rabitq, gao2025practical}) into (i) a 1-bit-per-dimension binary code kept in memory for fast approximate distances and (ii) an extended code (\eg 4 bits per dimension) for accurate refinement.
Prior work~\cite{gao2025practical} shows that a 1-bit binary plus 4-bit extended encoding balances search accuracy and storage footprint; we adopt this default.
Adjacency lists are sorted and integer-compressed (\eg delta encoding or Partitioned Elias-Fano~\cite{ottaviano2014partitioned}) to reduce space consumption.
The resulting variable-size record (\ie extended code plus compressed variable-size adjacency list) achieves an effective trade-off among storage footprint, search performance, and accuracy.

\begin{figure}[t]
    \centering
    \includegraphics[width=0.48 \textwidth]{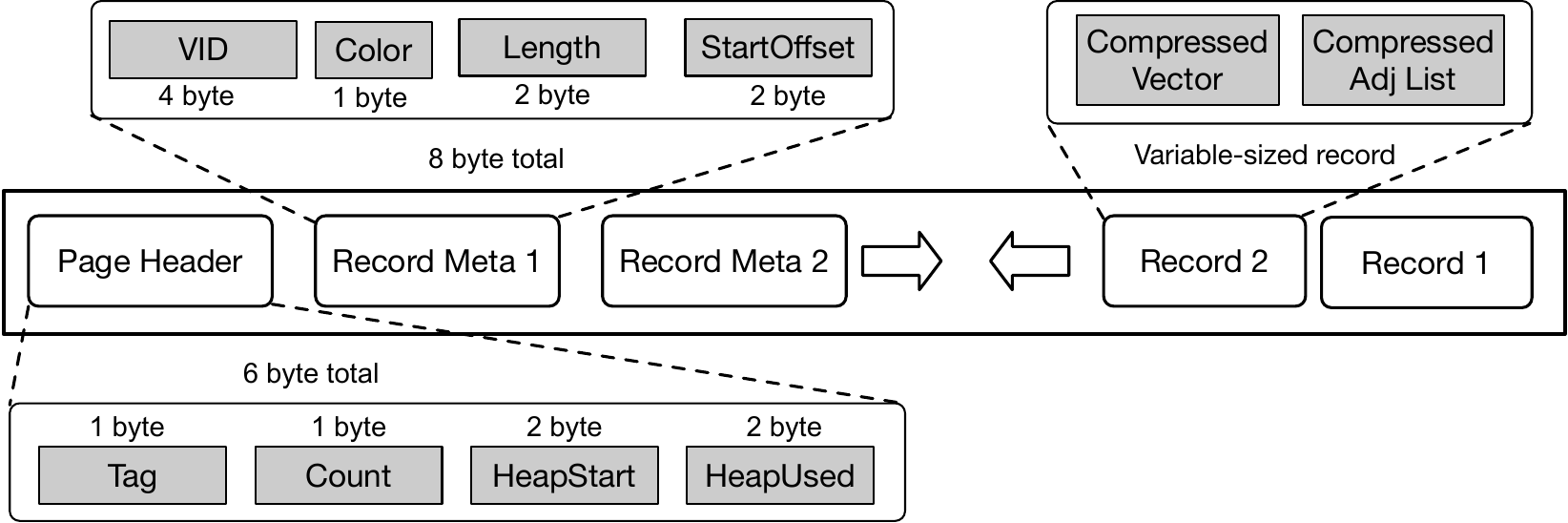}
    \caption{On-disk paging layout for organizing compressed variable-sized vertex-based records. }
    \Description{}
    \label{fig:page_layout}
\end{figure}

\para{Page Layout.}
To efficiently manage variable-size records on disk, we employ a B-tree-inspired slotted page layout, as illustrated in \fig\ref{fig:page_layout}. Each page is partitioned into a header, a slot array, and a data heap. The 5-byte \texttt{page header} provides page-management metadata, including a slot count (\texttt{Count}) and heap pointers (\texttt{HeapStart}, \texttt{HeapUsed}). The \texttt{slot array} of fixed-size entries grows from the start of the page, while the \texttt{data heap} storing records grows backward from the end; this opposing growth enables dynamic space sharing and minimizes internal fragmentation. Each 9-byte slot contains the record's 4-byte vertex ID (\texttt{VID}), a 1-byte \texttt{Color} tag to differentiate vertex clusters (detailed in \S\ref{sec:co_placement}), the record's 2-byte \texttt{Length}, and its 2-byte \texttt{StartOffset} within the heap. The slot array is kept sorted by \texttt{VID} to enable fast binary-search lookups. This design achieves dense packing of variable-size records and fast retrieval.
The retrieval process for an on-disk vertex record begins with a binary search on the sorted slot array to locate the metadata entry for a given vertex ID $v$.
The {\tt StartOffset} and {\tt Length} fields in this entry pinpoint the record's location and size within the data heap.
The system then reads the complete byte sequence, which constitutes the vertex-based record, and uses it for subsequent distance refinement and graph traversal.

\subsection{Record Co-Placement Mechanism}\label{sec:co_placement}
Record-level buffering improves cache utilization by loading only the requested record, but may introduce I/O inefficiency because page co-resident data is not retained.
Consider a traversal path $\{v_1 \!\rightarrow\! v_2 \!\rightarrow\! v_3\}$ with $v_1$ and $v_3$ on page $P_1$ and $v_2$ on $P_2$, yielding the page access sequence $\{P_1 \!\rightarrow\! P_2 \!\rightarrow\! P_1\}$.
Fetching $P_1$ for $v_1$ reads $v_3$ but discards it; when $v_3$ is later needed, $P_1$ is fetched again, incurring redundant disk I/O.
To mitigate this, \system uses affinity-based record co-placement to improve page-level locality, loading affine records together while discarding non-affine ones.

\para{Affinity-based Record Identification.}
Unlike block shuffling~\cite{wang2024starling}, which co-locates records of adjacent vertices within the same page, our approach exploits the empirical observation that spatially proximate vectors are frequently co-traversed along search trajectories.
This distinction is critical because modern proximity graphs (\eg Vamana~\cite{jayaram2019diskann}) optimize navigational efficiency rather than strictly preserving metric neighborhoods, employing long-range links and aggressively pruning edges among nearby vertices.
Consequently, adjacency-based co-location tends to pollute pages with records that are geometrically distant, degrading locality.
Formally, let $d(\cdot,\cdot)$ denote the distance metric. Two vectors $v_i$ and $v_j$ are deemed \emph{affine} iff $d(v_i,v_j)\!\le\!\tau$ for a given threshold $\tau$.
We co-locate records whose vectors are affine on the same page.
This strategy enables a single page read to serve multiple subsequent accesses, thereby reducing I/O operations.
The distance threshold $\tau$ is selected heuristically from dataset statistics.
We determine this threshold through vector clustering, setting $\tau$ to the average of the 5th-percentile distance-to-centroid values across all clusters.
This procedure can be integrated into the vector quantization stage, as many vector compression methods (\eg PQ~\cite{jegou2010product} or RabitQ~\cite{gao2024rabitq}) require a clustering step.

However, naively identifying affine records in a graph with $|P|$ vertices requires $O(|P|^2)$ pairwise checks, which is prohibitive at scale.
To address this, we integrate affinity identification into greedy proximity-graph construction with negligible additional overhead by reusing the candidate sets and distance computations produced during neighbor selection.
As shown in Algorithm~\ref{alg:affine} (lines 6--10), for each vertex $p$ we select up to $k$ nearest candidates within the threshold $\tau$ and store them in $S[p]$, avoiding an additional pass over the dataset.
Because the distance list $D$ is already computed for neighbor selection, affinity extraction reduces to per-vertex filtering.
We set the affinity bound $k$ relative to page capacity (\ie the maximum number of records per page) to prevent affinity groups from spanning multiple pages.
\begin{algorithm}[t]
\caption{Affine-based Record Identification}
\label{alg:affine}
\begin{algorithmic}[1]
\State \textbf{Input}: graph $G(P,E)$, build parameter $l_b$, graph degree $R$, pruning parameter $\alpha$, affinity threshold $\tau$, affine size $k$
\State \textbf{Output}: Graph $G(P',E')$ and affine record dictionary $S$
\State $S\gets \{\}$ \Comment{\textit{Initialize empty dictionary}}
\For {\textbf{each} $p$ in $P$}\Comment{\textit{Graph construction process}}
\State $[V,D]\gets$ \textsf{GreedySearch}($G$, $p$, $l_b$)
\begin{tcolorbox}[colback=gray!10,colframe=gray!10,left=-0.5mm,right=0mm,top=0mm,bottom=0mm,boxsep=0mm]
\State $A_p \gets \emptyset$ \Comment{\textit{Affine vertices for node $p$}}
\For{\textbf{each} $(v, d)$ \text{in} $(V, D)$}
\If{$d\!\le\!\tau$ and $|A_p| < k$}
\State $A_p \gets A_p \cup \{v\}$
\EndIf
\EndFor
\State $S[p] \gets A_p$ \Comment{\textit{Store affine vertices for node $p$}}
\end{tcolorbox}
\State{\textit{$\rhd$ Pruning mechanisms in proximity graph construction}}
\State $N_{out}(p)\gets$ \textsf{PruneEdges}($G$, $p$, $V$, $R$, $\alpha$)
\For{\textbf{each} $v$ in $N_{out}(p)$}
\State $N_{out}(v)\gets N_{out}(v)\cup \{p\}$
\If{$|N_{out}(v)|>R$}
\State $N_{out}(v)\gets$ \textsf{RobustPrune}($G$, $v$, $N_{out}(v)$, $R$, $\alpha$)
\EndIf
\EndFor
\EndFor
\State \textbf{return} $G(P',E')$, $S$
\end{algorithmic}
\end{algorithm}

\para{Affinity-based Record Co-Placement.}
We co-locate the affine records by iterating over the affinity dictionary and placing the sets contiguously on disk.
Within each page, we tag record slots with \texttt{Color} (see \fig~\ref{fig:page_layout}) to distinguish sets: each set receives a unique non-zero value, incremented cyclically; 0 denotes non-affine records.
One byte suffices for this tag because pages typically host few sets.
Pages are filled greedily.
If a set does not fit in the remaining space, we first pad the residual space with non-affine records (\ie records not belonging to any affinity set).
If none are available, we split the set across page boundaries to avoid the complexity and overhead of more sophisticated placement algorithms.
Upon accessing any record with a non-zero \texttt{Color} tag, all co-tagged records on the page are proactively fetched into the buffer pool.

\section{Cache-Aware ANNS Procedure}\label{sec:beam_search}
\system employs a coroutine-based asynchronous execution model, overlapping computation with I/O to keep the CPU saturated.
It achieves QPS comparable to fully in-memory ANNS systems, but single-query end-to-end latency remains higher due to disk I/O.
This discrepancy arises from I/O stalls on the query's critical execution path: {\bf while the system as a whole is asynchronous, a single query's critical path is not}; its coroutine blocks until its own I/O completes, making disk access effectively synchronous from that query's perspective.

Closing the latency gap hinges on reducing I/O stalls on each query’s critical path by improving the cache hit rate for requested vertex records.
However, ANNS workloads exhibit irregular memory access and poor cache locality.
To address this, \system introduces two complementary optimizations: \emph{proactive prefetching} and \emph{cache-aware beam search}.
Proactive prefetching (\S\ref{sec:prefetching}) issues asynchronous I/O for promising candidates ahead of use.
Cache-aware beam search (\S\ref{sec:cache_search}) alters the best-first search strategy to opportunistically prioritize exploring candidates already in memory, even if they are slightly sub-optimal.
Together, these techniques minimize critical-path stalls and effectively hide I/O latency.

\subsection{Prefetching}\label{sec:prefetching}

Our prefetching strategy is based on a key insight into the behavior of best-first graph traversal: the vertices to be explored in the immediate future are highly likely to be among the top-ranked candidates in the current search frontier.
This predictability stems from the best-first search algorithm's iterative nature, where it consistently expands the search frontier by selecting the most promising vertices based on their proximity to the query.

To capitalize on this predictability, \system employs a prefetching mechanism that preloads the records of the top-$B$ candidates in each search iteration.
unlike hardware prefetching, which is limited to in-memory data, our prefetching approach targets on-disk records.
\system implements this via a dedicated prefetching coroutine that runs concurrently with the main search process. 
The prefetching procedure reuses the asynchronous loading methods of the buffer pool manager.
As illustrated in \fig\ref{fig:coroutine_execution}, the prefetching coroutine identifies the top-$B$ most promising candidates in each search iteration and issues asynchronous I/O requests to load their records into the buffer pool.
Consequently, when the search algorithm proceeds to explore these candidates, their data is likely already present in memory, significantly improving the cache hit ratio and minimizing stalls on the query's critical path.

\begin{algorithm}[t]
\caption{Cache-aware Beam Search in \system.}
\label{alg:beam_search}
\begin{algorithmic}[1]
    \Statex \textbf{Input:} query $q$, candidate list size $L$, beam width $B$.
    \Procedure{VeloSearch}{$q, L, B$}
        \State Candidate list $p \gets \text{entry points}$
        \State Explored set $E \gets \emptyset$
        \While{$P \not\subseteq E$}
            \State $v \gets$ top-1 nearest candidate to $q$ from $P \setminus E$
            \begin{tcolorbox}[colback=gray!10,colframe=gray!10,left=-0.5mm,right=0mm,top=0mm,bottom=0mm,boxsep=0mm]
            \State $C \gets \text{top-}B \text{ nearest candidates to} q \text{ from } P \setminus E$
            \State \Comment{{\it Prioritize the in-memory candidate when $v$ is on disk\,\,}}
            \If {InOnDisk($v$)}
                \For{\textbf{each} $c$ in $C$}
                    \If {InMemory($c$)}
                    \State $v \gets c$; \textbf{break}
                    \Else
                    \State prefetch($c$)\Comment{{\it Prefetch on-disk candidate $c$}}
                    \EndIf
                \EndFor
            \EndIf
            \end{tcolorbox}
            \State $v$ = read($v$){\bf .await}\Comment{\textit{Suspend execution if $v$ is on disk}}
            \For{\textbf{each} nbr in $v.\text{neighbors} \setminus E$}
                \State $dis \gets \text{estimate\_distance}(\text{nbr}, q)$
                \State $P.\text{insert}(\langle \text{nbr}, dis \rangle)$
            \EndFor
            \State $E.\text{insert}(v.\text{neighbors})$
            \State Trim candidate set $P$ to size $L$
        \EndWhile
        \State \textbf{return} top-$k$ nearest vectors to $q$ in $E$
    \EndProcedure
\end{algorithmic}
\end{algorithm}

\subsection{Cache-aware Beam Search}\label{sec:cache_search}
The best-first search procedure creates a strong dependency between I/O and computation, as the algorithm must wait for candidate records to be fetched from disk before proceeding.
To mitigate this, beam search, as proposed by DiskANN~\cite{jayaram2019diskann}, explores a batch of $B$ vertex candidates concurrently instead of a single nearest vertex.
However, this approach remains constrained by a sequential compute-I/O pattern and is ultimately bottlenecked by the slowest I/O operation in each batch.

To address this, we introduce a cache-aware beam search that leverages \system's efficient buffer pool.
Unlike standard beam search, which synchronously issues a batch of I/O requests, our approach prioritizes the exploration of candidates that are cached in memory, while asynchronously prefetching those on disk.
This strategy yields two key advantages.
First, it amplifies the effective cache hit rate by a factor of $B$, as the probability of finding an in-memory candidate increases proportionally with the beam size.
Second, it minimizes I/O stalls by prioritizing the exploration of in-memory candidates while asynchronously prefetching other candidates in the background, effectively hiding I/O latency.

Algorithm~\ref{alg:beam_search} illustrates our cache-aware search procedure with beam width $B$, which deviates from a standard greedy beam search by opportunistically prioritizing in-memory candidates.
In each iteration, the algorithm first identifies the top candidate $v$ from the candidate list $P$ (line~6) and constructs a look-ahead set $C$ containing the top-$B$ candidates (line~7).
If $v$ resides on disk (line~9), the algorithm attempts an opportunistic pivot by iterating through the look-ahead set $C$.
It selects the first available in-memory candidate $c$ for immediate exploration (lines~10--12), allowing computation to proceed without stalling.
For any on-disk candidates encountered in $C$, it issues asynchronous prefetch requests (line~14).
If no in-memory candidate is found within the look-ahead set, the search defaults to processing $v$, potentially suspending execution until its data is retrieved from disk (line~17).
After a vertex is explored, its unvisited neighbors are added to the candidate list, and the vertex is marked as explored (lines~18--22).
Finally, the candidate list is trimmed to a maximum size of $L$ (line~22).
This design effectively hides I/O latency by combining the opportunistic exploration of in-memory candidates with asynchronous data prefetching, thereby reducing I/O stalls and overall query latency.

\begin{figure*}[t]
    \centering
    \includegraphics[width=1\textwidth]{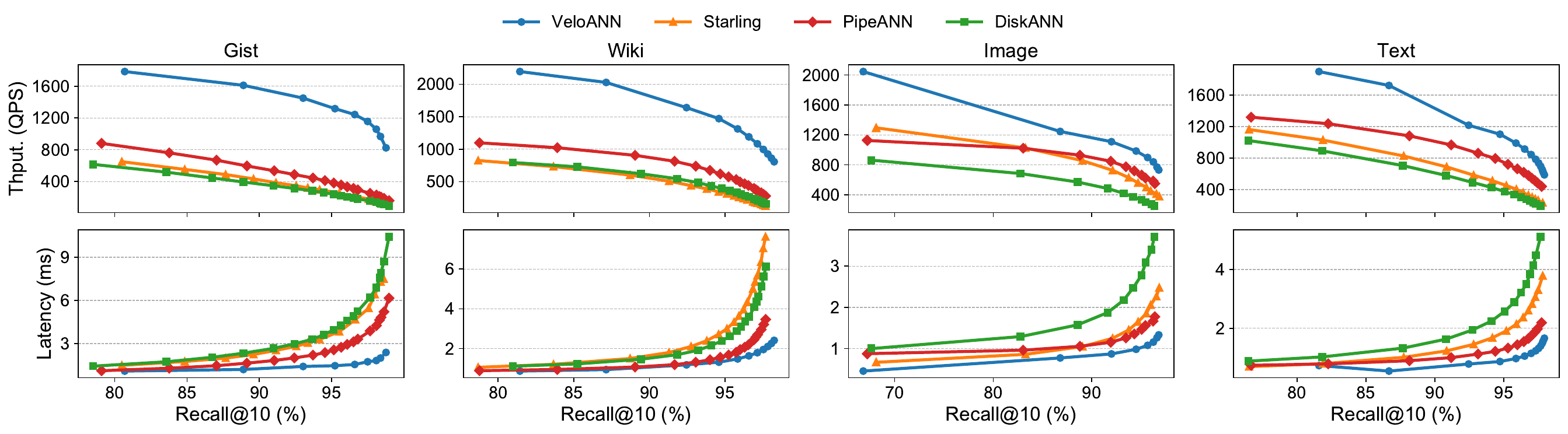}
    \caption{Performance comparison of \system with DiskANN, Starling, and PipeANN on four high-dimensional datasets.}
    \label{fig:performance_comparison}
\end{figure*}

\begin{table}[t]
\centering
\caption{Statistics of the datasets.}
\begin{tabular}{lcccc}
\hline
Dataset & \# of vectors & Dimension & Size (GB) & \# of queries \\
\hline
Sift1M & 1M & 128 & 0.52 & 10,000 \\
GIST1M & 1M & 960 & 1.28 & 1,000 \\
Wiki& 35M & 768 & 101 & 10,000 \\
Image & 100M & 512 & 191.7 & 10,000 \\
Text & 100M & 768 & 286.4 & 10,000 \\
\hline
\end{tabular}
\label{tab:datasets}
\end{table}

\section{Evaluation}
\label{sec:exp}
\subsection{System Implementation}
We implemented \system in $\sim$13k lines of Rust code, chosen for its powerful asynchronous programming features and strong safety guarantees.
The query engine of \system adopts a thread-per-core asynchronous execution model, where each thread runs a dedicated scheduler to minimize context-switching overhead and enhance cache locality.
For high-performance asynchronous I/O operations, we implement the asynchronous runtime based on {\tt io\_uring}, with each thread employing a private instance to issue I/O requests independently.
To reduce memory access overhead, the buffer manager allocates its memory region using {\tt mmap} with huge pages.
\system employs ExtRabitQ~\cite{gao2024rabitq,gao2025practical} for vector compression and the Partitioned Elias-Fano coding~\cite{ottaviano2014partitioned} for compressing adjacency lists, creating a compact graph layout that is tightly integrated into the query engine.
Rust's compile-time checks provide strong guarantees for memory and concurrency safety throughout the system.

\subsection{Experimental Setup}
{\bf Testbed.}
We conduct our evaluation on a server equipped with a 48-core Intel(R) Xeon(R) Platinum 8457C processor operating at 2.00GHz, 384GB of RAM configured as 24$\times$16GB DDR4 modules running at 2933MT/s, and two 3.84TB Solidigm NVMe SSDs. The system runs Ubuntu 24.04 LTS with Linux kernel version 6.8.0.

\noindent{\bf Datasets.} Our evaluation employs five representative benchmark datasets spanning diverse dimensions and scales, as detailed in Table~\ref{tab:datasets}.
These include two commonly used Sift1M~\cite{jegou2011searching} and Gist1M~\cite{jegou2011searching} datasets, for which we utilize the provided base and query sets for micro-level analysis.
Additionally, we incorporate three datasets constructed from high-dimensional embeddings generated by advanced deep learning models. 
The first is a 512-dimensional text dataset created using the Cohere~\cite{Cohere} embedding model, with 35M base vectors generated from Wikipedia and 10K query vectors sampled from the MMLU~\cite{hendrycks2020measuring} dataset.
The second and third datasets are 100M-scale high-dimensional embeddings. Image consists of 512-dimensional image embeddings with a 100M-vector base set and 10K query vectors, while Text comprises 768-dimensional text embeddings with a 100M-vector base set and 10K query vectors.

\noindent{\bf Compared Systems.}
We benchmark \system against three state-of-the-art disk-based ANNS systems: DiskANN~\cite{jayaram2019diskann}, Starling~\cite{wang2024starling}, and PipeANN~\cite{guo2025achieving}.
DiskANN serves as a widely adopted baseline for disk-based ANNS system.
Starling enhances search locality through on-disk graph layout reordering and employs an in-memory index to reduce I/O traversal paths.
PipeANN accelerates on-disk search by relaxing strict best-first ordering to parallelize I/O requests and search procedures.
To ensure a fair comparison, we standardize the experimental setup by using RabitQ~\cite{gao2024rabitq} for in-memory vector compression across all systems and replacing AIO with {\tt io\_uring} in DiskANN and Starling.
We enable the {\tt O\_DIRECT} option for all systems to bypass the operating system page cache.
Since all systems utilize the Vamana~\cite{jayaram2019diskann} graph index, we employ identical construction hyperparameters when building the graphs.

To ensure a fair comparison, we set the buffer ratio ({\ie} memory budget) for each system to 20\% of its respective disk index size.
This normalization prevents \system, which has smaller on-disk compressed indexes, from receiving a disproportionately large buffer pool allocation.
We disable the {\tt SQ\_POLL} feature of {\tt io\_uring} because it introduces an additional thread to busy-poll I/O completion events, which would compromise fairness in the comparsion.
Furthermore, we observe that for high-dimensional datasets (\eg Gist1M with 960 dimensions), individual vertex records can occupy an entire data page (default 4\,KB).
To address this issue, we configure baseline systems with an 8\,KB page size for these datasets, while \system employs a 4\,KB page size.

\noindent{\bf Evaluation Metrics.} We evaluate and report the following key metrics: Recall@10 represents the fraction of the true 10 nearest neighbors retrieved in the top-10 results, measuring the quality of the search results.
Query latency indicates the average per-query processing time in millisecond (ms).
Throughput refers to the number of queries processed per second under concurrent load (QPS).
The number of disk I/Os indicates the average number of I/O requests issued per query.

\subsection{Overall Performance}
\para{Query Throughput and Latency.}
As illustrated in \fig\ref{fig:performance_comparison}, we compare the query throughput and latency of \system against three baselines across four high-dimensional datasets.
The results demonstrate that \system consistently outperforms all baselines in both query throughput and latency at the same recall@10.
\system's asynchronous execution model orchestrates multiple query requests to overlap I/O and computation, yielding significant throughput gains.
This results in a query throughput 1.4$\times$ to 4.6$\times$ higher than PipeANN.
Furthermore, by leveraging prefetching and a cache-aware beam search, \system effectively mitigates I/O latency from the critical execution path, achieving 1.6$\times$ lower latency than the latency-optimized PipeANN at high recall.
For instance, on the Wikipedia dataset, \system delivers over 1k QPS at 97\% recall, surpassing PipeANN, Starling, and DiskANN by 3.3$\times$, 7.3$\times$, and 5.8$\times$, respectively. 
In terms of query latency, \system also achieves 1.44$\times$, 3.25$\times$ and 2.5$\times$ lower latency than DiskANN, Starling, and PipeANN, respectively.
We also observe that Starling's performance is comparable to DiskANN's, despite its larger 8\,KB page size setting.
This is because an 8\,KB page accommodates only a few high-dimensional records (\eg three 512D or one 768D vector), which diminishes the benefit of its block shuffling layout and underscores the importance of a compressed index for high-dimensional ANNS.

\begin{table}[t]
\centering
\small
\setlength{\tabcolsep}{3pt}
\caption{Comparison of index size and memory footprint across different ANN systems. The \textit{Origin} row shows the size of the raw vector datasets for reference.}
\label{tab:index_size}
\begin{tabular}{@{}lcccc|cccc@{}}
\toprule
\multirow{2}{*}{\textbf{}} & \multicolumn{4}{c}{\textbf{Disk (GB)}} & \multicolumn{4}{c}{\textbf{Memory Footprint (GB)}} \\
\cmidrule(lr){2-5} \cmidrule(lr){6-9}
& Gist & Wiki & Image & Text & Gist & Wiki & Image & Text \\
\midrule
{\it Origin} & 4.2 & 96 & 286 & 201 & -- & -- & -- & -- \\
DiskANN & 8.0 & 135 & 382 & 286 & 0.87 & 24.5 & 49.7 & 38.6 \\
VeloANN & \textbf{0.8} & \textbf{21} & \textbf{64} & \textbf{45} & \textbf{0.30} & \textbf{7.3} & \textbf{20.8} & \textbf{15.7} \\
\bottomrule
\end{tabular}
\end{table}

\para{Index Size and Memory Footprint.}
As the three baseline systems share the same underlying storage layout (differing only in ordering for Starling), we compare the index size and memory footprint of \system with those of DiskANN as illustrated in Table~\ref{tab:index_size}.
The result shows that \system's disk consumption is up to 10$\times$ smaller than DiskANN.
This significant space saving is achieved through a compressed index layout that quantizes high-dimensional vectors to 4 bits per dimension and compresses each vertex's adjacency list with a sorted integer encoding scheme.
These optimizations yield a nearly 4.5$\times$ compression ratio compared to the original raw vectors (labeled as {\it Origin}).
In contrast, baseline systems exhibit significant space amplification.
This is because, in addition to graph structure overhead, the DiskANN-style layout suffers from severe internal page fragmentation especially for high-dimensional vector data.
For instance, on the Image dataset, a standard 4\,KB page accommodates only one record---a 2,048-byte vector and a 256-byte adjacency list for 64 out-degrees---wasting 1,792 bytes per page.
By mitigating this fragmentation, \system's compact encoding scheme enables multiple records to be stored per page, demonstrating its effectiveness for managing high-dimensional datasets.

As detailed in Table~\ref{tab:index_size}, \system also achieves the lowest memory footprint, requiring up to 2.9$\times$ less memory than that of DiskANN.
This footprint comprises three components: highly-compressed (1-bit per dimension) vectors for initial distance estimation, a buffer pool for caching vertex records, and auxiliary metadata for the buffer manager.
While we set the buffer pool to 20\% of the on-disk index size in our experiments, this ratio is a flexible parameter for tuning performance and memory trade-offs.
The metadata, which includes record mappings and states, adds only a minor overhead of approximately 1\% of the disk index size.

\subsection{Micro Analysis}
\begin{figure}[t]
    \centering
    \includegraphics[width=0.46\textwidth]{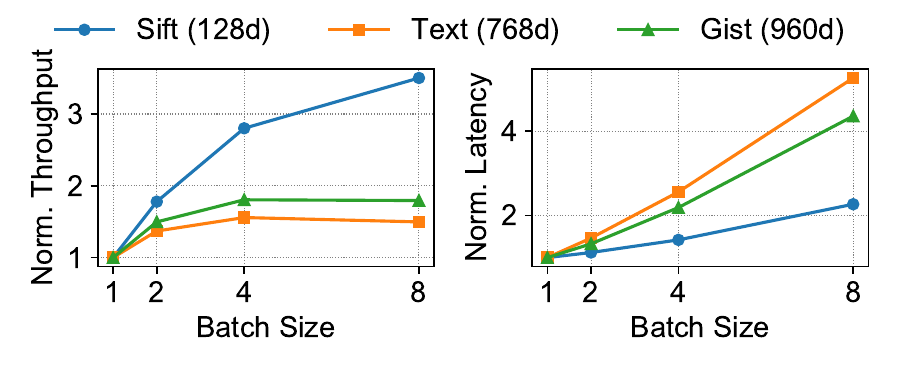}
    \caption{Query throughput and latency of \system with varying batch size (B) settings on three datasets.}
    \label{fig:batch_size}
\end{figure}

\para{Batch Size.} 
As described in \S\ref{sec:coroutine_execution}, the scheduler's batch size ($B$) creates a trade-off between throughput and latency.
A larger batch size can improve throughput by more effectively hiding I/O latency, as the scheduler can execute other queries during I/O stalls.
Conversely, a larger batch may increase average query latency due to computational and scheduling overhead.
\fig\ref{fig:batch_size} illustrates this behavior, showing that query latency grows approximately linearly as $B$ increases.
The impact on throughput, however, is dataset-dependent.
On the lower-dimensional Sift1M dataset, throughput scales effectively with larger batch sizes.
This is because its low computational cost allows the scheduler to mask I/O wait times by executing more concurrent queries.
In contrast, on the high-dimensional Gist1M and Image datasets, throughput gains diminish beyond $B=2$.
For these datasets, the high computational cost saturates the CPU even with a small batch, making larger batches less effective.
Based on these findings, we recommend a default batch size of $B=2$ for high-dimensional datasets, as it provides a strong balance between throughput and latency.

\begin{figure}[t]
    \centering
    \includegraphics[width=0.46\textwidth]{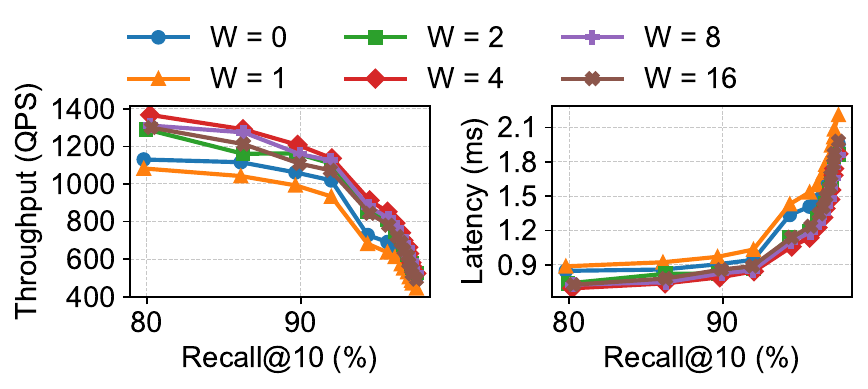}
    \caption{Query throughput and latency of \system under different beam width (W) settings on the Image dataset.}
    \label{fig:beam_size}
\end{figure}

\para{Beam Width.}
The beam width ($W$) in our cache-aware ANNS procedure presents a trade-off between I/O latency hiding and I/O efficiency.
Increasing the beam width expands the set of candidates for exploration, which improves the likelihood of cache hits and allows the scheduler to hide I/O latency by processing other candidates.
However, this also leads to wasted I/O bandwidth compared to a best-first search and extra prefetching coroutine overhead, as the algorithm can prefetch suboptimal vertices that are ultimately pruned from the final search path.
\fig\ref{fig:beam_size} illustrates this trade-off on the Image dataset, showing how query throughput and latency vary with different beam widths.
As shown, performance peaks at $W=4$, after which it begins to decline.
Interestingly, a beam width of $W=1$ results in worse performance than a standard best-first search ($W=0$).
This occurs because the time required to prefetch a single vertex from disk is greater than the processing time for an in-memory vertex.
Consequently, the search process frequently stalls while waiting for the I/O to complete, rendering the single-vertex prefetching ineffective.
Therefore, selecting an appropriate beam width is crucial for achieving optimal performance.

\begin{figure}[t]
    \centering
    \includegraphics[width=0.46\textwidth]{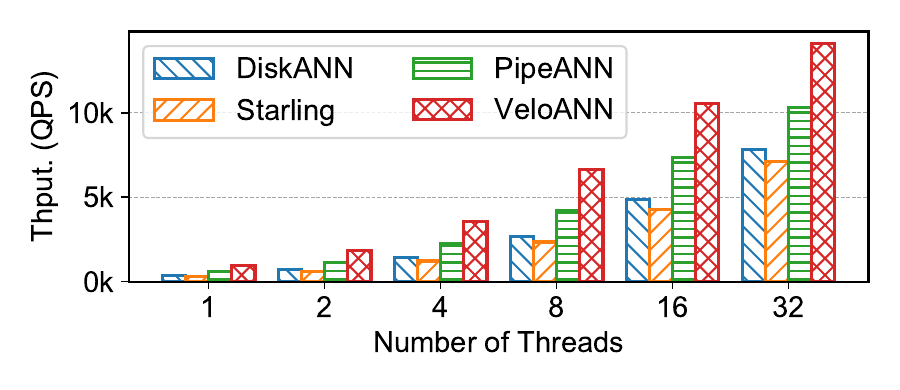}
    \caption{Query throughput of DiskANN, Staring, PipeANN and \system with varying thread number settings on the Wikipedia dataset.}
    \label{fig:threads}
\end{figure}

\para{Varying Thread Number.}
\fig\ref{fig:threads} evaluates the throughput scalability of \system against its baselines as the number of concurrent worker threads increases from 1 to 32.
Each system is configured to achieve approximately 95\% recall@10.
We employ a round-robin strategy to distribute queries evenly among the threads.
\system demonstrates superior scalability, consistently outperforming all baselines by a significant margin.
\system employs an effective record-level buffer pool to reduce disk I/Os and leverages coroutine-based execution to enhance query throughput.
At 32 threads, \system reaches a peak throughput of $\sim$14k QPS.
This represents a 1.39$\times$ speedup over PipeANN and more than a 1.82$\times$ speedup over DiskANN and Starling.
This superior scalability demonstrates the \system's I/O efficiency.

\begin{figure}[t]
    \centering
    \includegraphics[width=0.46\textwidth]{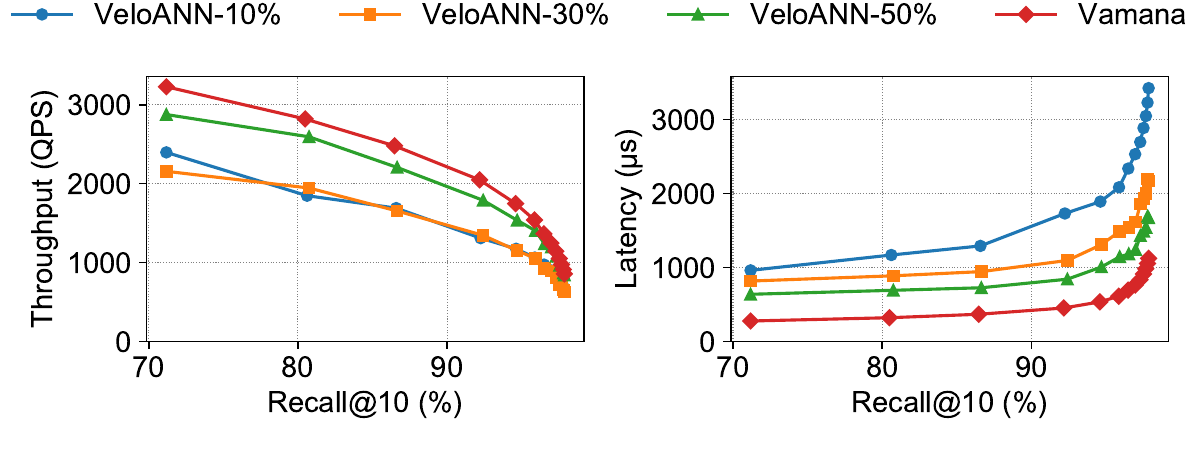}
    \caption{Query throughput and latency of \system on the Image dataset with varying buffer ratios, compared to a fully in-memory Vamana index.}
    \label{fig:in_memory}
\end{figure}

\para{Comparing to In-Memory ANNS Indexes.}
\fig\ref{fig:in_memory} illustrates the performance of \system on the Image dataset with varying buffer ratios, compared to a fully in-memory Vamana index (\ie a 100\% buffer ratio).
By leveraging its asynchronous execution model, \system effectively saturates the CPU with computational tasks, avoiding idle periods typically spent waiting for I/O completion.
Furthermore, its prefetching and cache-aware beam search strategies significantly reduce I/O stalls in the critical query path, effectively hiding much of the disk I/O latency.
As a result of these optimizations, \system achieves throughput comparable to the in-memory Vamana index.
Specifically, at 95\% Recall@10, it delivers 0.73$\times$, 0.78$\times$, and 0.92$\times$ the QPS of Vamana with buffer ratios of 10\%, 30\%, and 50\%, respectively.
Correspondingly, at the same recall target, the query latencies are 2.23$\times$, 2.19$\times$, and 1.86$\times$ higher than Vamana.
These results demonstrate that \system offers a cost-effective trade-off, achieving, for instance, 78\% of in-memory Vamana's throughput while using only 30\% of the index size in memory (equivalent to 7\% of the original dataset size).

\begin{figure}[t]
    \centering
    \includegraphics[width=0.46\textwidth]{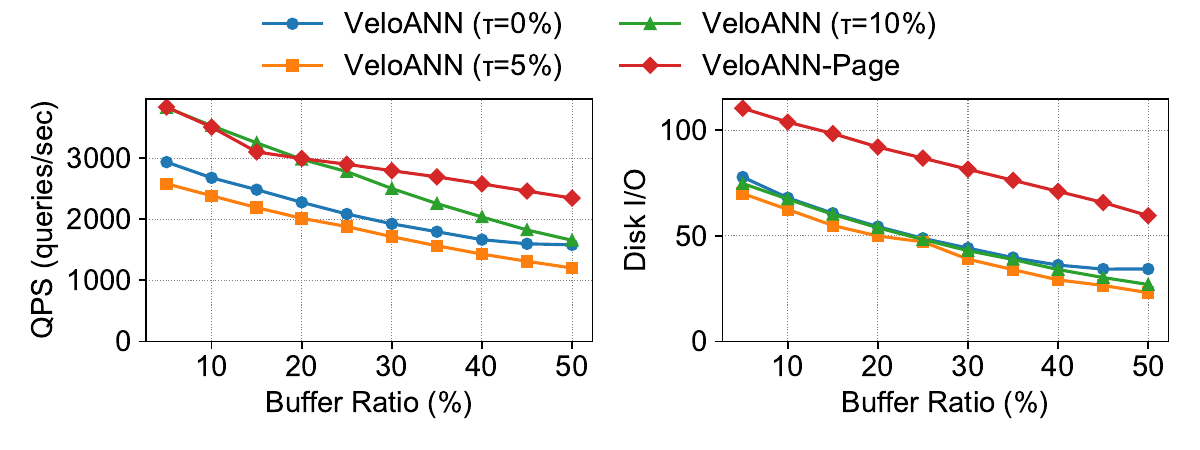}
    \caption{Query latency and disk I/O count under different distance threshold $\tau$ settings on the Sift1M dataset.}
    \label{fig:record_cache}
\end{figure}

\para{Record Co-Placement with Varying Threshold $\tau$.}
The distance threshold $\tau$ is crucial for record co-placement mechanism.
We evaluate its effect by comparing VeloANN with different $\tau$ settings against VeloANN-Page, a baseline that loads all records into the buffer pool from a page.
As illustrated in \fig\ref{fig:record_cache}, the page-based approach (VeloANN-Page) performs the worst, exhibiting the highest disk I/O count and query latency.
This is because loading entire pages pollutes the buffer with unnecessary records, evicting more valuable data.
In contrast, VeloANN with $\tau=0\%$ fetches only a single record, avoiding pollution but missing I/O consolidation opportunities.
Setting $\tau=5\%$ strikes an optimal balance, achieving the lowest query latency and disk I/O, up to 1.28$\times$ lower than the $\tau=0\%$ case.
However, increasing the threshold further to $\tau=10\%$ degrades performance.
A larger $\tau$ relaxes the affinity requirement for co-placement, reducing the effectiveness of I/O consolidation and leading to worse performance.

\subsection{Breakdown Analysis}
In this section, we analyze the performance gap between Baseline and \system by progressively enabling the proposed optimizations.
Based on the index layout shuffled by record co-placement mechanism, we incrementally integrate the key techniques into Baseline and evaluate query throughput and latency across target recall levels.
We fix the memory buffer ratio at 10\% in bellow experiments.

\para{+Async.}
To address CPU underutilization, we introduce an asynchronous execution model (\S\ref{sec:coroutine_execution}) for Baseline.
This variant overlaps I/O with computation (\eg distance comparison operations) to maximize CPU utilization, yielding 1.8$\times$ higher throughput at 95\% recall@10 than Baseline.
However, frequent context switches among query coroutines increase per-query latency, about 1.6$\times$ higher at 95\% recall@10.

\begin{figure}[t]
    \centering
    \includegraphics[width=0.46\textwidth]{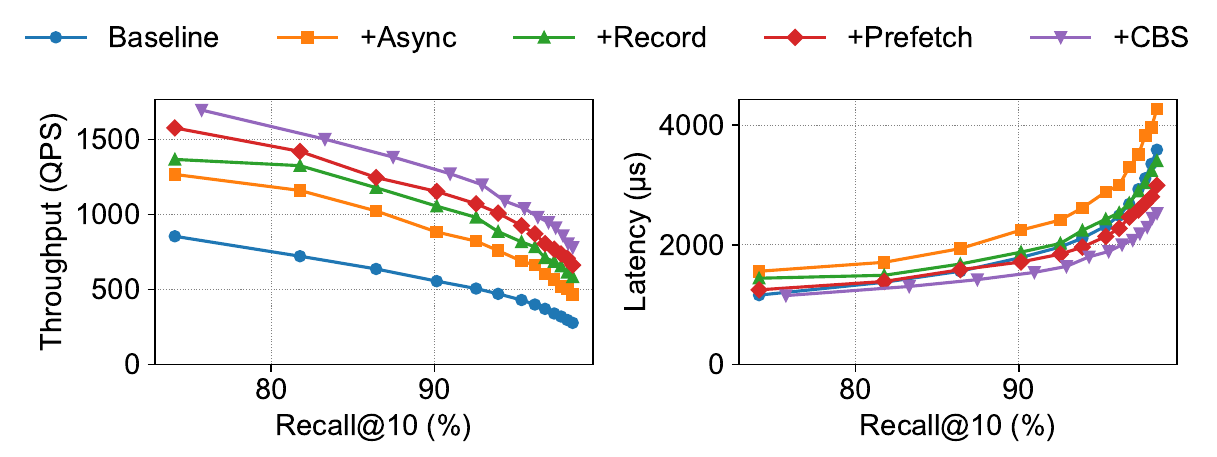}
    \caption{Breakdown analysis of \system on Text dataset.}
    \label{fig:breakdown}
\end{figure}

\para{+Record.}
We integrate a record-level buffer pool (\S\ref{sec:buffer_pool}) that provides finer-grained memory management compared to page-level caching.
This optimization yields a query throughput improvement of 1.23$\times$ to 1.9$\times$ over the async variant and Baseline, respectively.
Additionally, it achieves a 1.21$\times$ reduction in query latency relative to the async variant.
These performance gains primarily result from reduced disk I/O operations enabled by record-level caching.
The fine-grained caching mechanism avoids loading unnecessary records into memory, thereby preserving memory resources for more frequently accessed record data.

\para{+Prefetch.}
To further mitigate disk I/O stalls caused by cache misses, we introduce a prefetching strategy (\S\ref{sec:beam_search}) that performs stride prefetching of records for the next several vertices in the candidate list.
By anticipating future record requests based on the current candidate list, this strategy proactively fetches the corresponding records into the buffer pool.
This approach significantly improves the cache hit rate and reduces I/O stalls in the critical query path.
As shown in \fig\ref{fig:breakdown}, this method enhances query throughput by 2.15$\times$ at 95\% recall@10 and decreases average query latency by 1.34$\times$ compared to the async variant.

\para{+CBS (Cache-Aware Beam Search).}
Integrating the above optimizations, \system further reduces latency with cache-aware beam search.
CBS opportunistically prioritizes in-memory candidates within the beam, increasing cache hit rate and reducing critical-path I/O stalls.
At 95\% recall@10, CBS attains the lowest latency, 1.5$\times$ lower than the async variant, and improves throughput by 1.5$\times$ over the async variant and 2.2$\times$ over Baseline.
In terms of query throughput, it achieves a 1.5$\times$ improvement over the async variant and a 2.2$\times$ improvement over the baseline at 95\% recall@10.

\section{Related Work}
\label{sec:relatedwork}

\para{ANNS index.}
Research on ANNS indexes is generally categorized into three types: cluster-based~\cite{chen2021spann,xu2023spfresh,andre2016cache,dong2019learning}, hashing-based~\cite{tian2023db, zhao2023towards,liu2014sk,datar2004locality,gan2012locality}, and graph-based~\cite{malkov2018efficient,harwood2016fanng,fu2017fast,wang2021comprehensive,azizi2023elpis,peng2023efficient} indexes.
Among these, graph-based approaches stand out in large-scale, high-dimensional search by offering both high speed and accuracy.
However, as dataset sizes continue to grow, the memory overhead of in-memory graph-based ANNS becomes a significant bottleneck, limiting their scalability and applicability.

\para{On-Disk ANNS Storage.}
To cost-effectively support large-scale vector datasets, recent researches have focused on storing ANNS indexes on disk.
Cluster-based methods, such as SPANN~\cite{chen2021spann} and ScaNN~\cite{guo2020accelerating}, partition vectors into clusters that are stored on disk and indexed by an in-memory graph.
DiskANN~\cite{jayaram2019diskann}, a pioneering on-disk graph-based system, stores compressed vectors in memory for I/O-free navigation and retrieves full-precision vectors from disk only to refine search results.
Subsequent works like DiskANN++~\cite{ni2023diskann++} and Starling~\cite{wang2024starling} introduce further optimizations, including sampled in-memory indexes to shorten I/O traversal paths and topology-guided data co-placement to enhance locality.
Others focus on layout optimization; for instance, Gorgeous~\cite{yin2025gorgeous} introduces a memory cache that prioritizes the graph structure over vector data, as the former is accessed more frequently.
To improve update performance, OdinANN~\cite{guo2026odin} utilizes direct insertion into the on-disk index, avoiding the costly batch-merge process of other systems.
In contrast, \system achieves superior I/O efficiency through a combination of a compressed on-disk graph layout, memory-optimized record caching, and an integrated layout shuffling process during graph construction that incurs negligible additional overhead.

\para{I/O-Efficient ANNS Query Processing.} 
To enhance I/O efficiency in query processing, cluster-based index SPANN identify the nearest clusters in memory and utilize batched I/O to amortize latency.
However, their coarse-grained nature, compared to graph-based indexes, often results in lower search throughput.
DiskANN~\cite{jayaram2019diskann} pioneered a beam search strategy that concurrently fetches multiple candidate vectors at each step to mitigate I/O stalls.
This approach, however, is constrained by a sequential compute-I/O pattern, where performance is bottlenecked by the slowest I/O operation in each batch.
To improve the I/O-efficiency, Starling~\cite{wang2024starling} proposes a block search method that explores additional vertices within the same data block for each disk I/O, exploiting the data locality from its shuffling strategy.
PipeANN~\cite{guo2025achieving} further improves upon this by pipelining computation and I/O, aligning its best-first search with modern SSD characteristics to enable effective overlap.
In contrast, \system introduces a fully asynchronous search procedure coupled with a record-level memory cache, which minimizes I/O stalls on the critical path and achieves performance comparable even to fully in-memory indexes.
\section{Conclusion}
\label{sec:conclusion}
We present \system, a graph-based ANNS for workloads exceeding main-memory capacity.
It couples an optimized on-disk layout with a coroutine-based asynchronous runtime to reduce storage stalls and improve CPU utilization.
Built on a record-level buffer pool, \system integrates proactive prefetching and cache-aware beam search to raise cache hit rates and overall performance.
Experiments show that \system outperforms state-of-the-art disk-based baselines in throughput and latency, while approaching in-memory performance with a smaller memory footprint.



\bibliographystyle{ACM-Reference-Format}
\bibliography{sample}

\end{document}